\documentclass[sigconf,10pt]{acmart}

\AtBeginDocument{%
  }

\usepackage{algorithmic}
\usepackage{graphicx}
\usepackage{textcomp}
\usepackage{xcolor}
\usepackage{comment}
\usepackage{listings}
\usepackage{soul}
\usepackage{xurl}
\usepackage{todonotes}
\usepackage{pifont}
\usepackage{adjustbox}
\usepackage{array}
\usepackage{multirow}
\usepackage{booktabs}
\usepackage{tikz}
\definecolor{codegreen}{RGB}{0, 155, 85}

\lstdefinestyle{mystyle}{
    backgroundcolor=\color{white},   
    commentstyle=\color{blue},
    otherkeywords = {:,[,],\$,\{,\},(,)},
    numberstyle=\tiny\color{black},
    stringstyle=\color{black},
    basicstyle=\ttfamily,
    breakatwhitespace=false,         
    breaklines=true,                 
    captionpos=b,                    
    keepspaces=true,                 
    numbers=left,                    
    numbersep=5pt,                  
    showspaces=false,                
    showstringspaces=false,
    showtabs=false,                
    }
\newcommand{\DrawPercentageBar}[1]{%
  \begin{tikzpicture}
    \fill[color=black]   (0.0 , 0.0) rectangle (#1*3ex , 1.5ex );
    \fill[color=gray] (#1*3ex  , 0.0) rectangle (3.0ex, 1.5ex);
  \end{tikzpicture}%
}
\newcommand{\DrawPercentageBarBlue}[1]{%
  \begin{tikzpicture}
    \fill[color=blue]   (0.0 , 0.0) rectangle (#1*3ex , 1.5ex );
    \fill[color=gray] (#1*3ex  , 0.0) rectangle (3.0ex, 1.5ex);
  \end{tikzpicture}%
}

\setcopyright{acmlicensed}
\copyrightyear{2025}
\acmYear{2025}
\acmDOI{XXXXXXX.XXXXXXX}

\acmBooktitle{The 18th European Workshop on Systems Security (EuroSec '25),
  March 31, 2025, Rotterdam, Netherlands}
\acmISBN{978-1-4503-XXXX-X/18/06}

\begin{document}
\acmYear{2025}\copyrightyear{2025}
\acmConference[EuroSec '25]{The 18th European Workshop on Systems Security}{March 30--April 3, 2025}{Rotterdam, Netherlands}
\acmBooktitle{The 18th European Workshop on Systems Security (EuroSec '25), March 30--April 3, 2025, Rotterdam, Netherlands}
\acmDOI{10.1145/3722041.3723097}
\acmISBN{979-8-4007-1563-1/25/03}
\title[Can Neural Decompilation Assist Vulnerability Prediction on Binary Code?]{Can Neural Decompilation Assist\\ Vulnerability Prediction on Binary Code?}
\author{Domenico Cotroneo, Francesco C. Grasso, Roberto Natella, Vittorio Orbinato}

\email{{cotroneo, francescocrescenzo.grasso, roberto.natella, vittorio.orbinato}@unina.it}

\affiliation{%
  \institution{Università degli Studi di Napoli Federico II}
  \city{Naples}
  \state{}
  \country{Italy}
}

\begin{abstract}
Vulnerability prediction is valuable in identifying security issues efficiently, even though it requires the source code of the target software system, which is a restrictive hypothesis. This paper presents an experimental study to predict vulnerabilities in binary code without source code or complex representations of the binary, leveraging the pivotal idea of decompiling the binary file through \textit{neural decompilation} and predicting vulnerabilities through deep learning on the decompiled source code. The results outperform the state-of-the-art in both neural decompilation and vulnerability prediction, showing that it is possible to identify vulnerable programs with this approach concerning bi-class (vulnerable/non-vulnerable) and multi-class (type of vulnerability) analysis.
\end{abstract}




\maketitle

\section{Introduction}
\label{sec:introduction}
\textit{Vulnerability prediction} is an approach to analyze and predict which files or functions of a software system are most likely vulnerable. This approach enables focusing effort and resources (e.g., man-hours) on the vulnerable parts. 
However, vulnerability prediction has the drawback of being a source code-based approach, which is a restrictive hypothesis: source code is rarely available for several classes of software systems, such as proprietary software, legacy systems, or firmware monitoring specific hardware. Limited access to source code hinders vulnerability prediction, hence the need for new solutions to identify vulnerabilities when source code is unavailable \cite{yan2021han, yamaguchi2012generalized, xu2018vulnerability}.

\textit{Neural decompilation} is an emerging method that leverages modern deep learning techniques, such as transformers and self-attention, to translate binary code back into source code. One of the key benefits of this approach is that the control flow structure of the decompiled program resembles the original source code, in a closer way than disassemblers and traditional decompiler tools, such as Ghidra \cite{Ghidra}. 
Neural decompilation is a data-driven approach, where the neural network learns the typical patterns of source code structures. Gaining access to the source code offers several advantages, such as enabling vulnerability detection, malware recognition, re-engineering of legacy systems, and facilitating static analysis. Moreover, binary decompilation enables vulnerability prediction in ``black-box'' vulnerability assessments (e.g., proprietary software), where the assessor attacks the system from the same perspective as the attacker, without source code.

\begin{figure*}[!htbp]
\centering
\centerline{\includegraphics[width=0.7\textwidth]{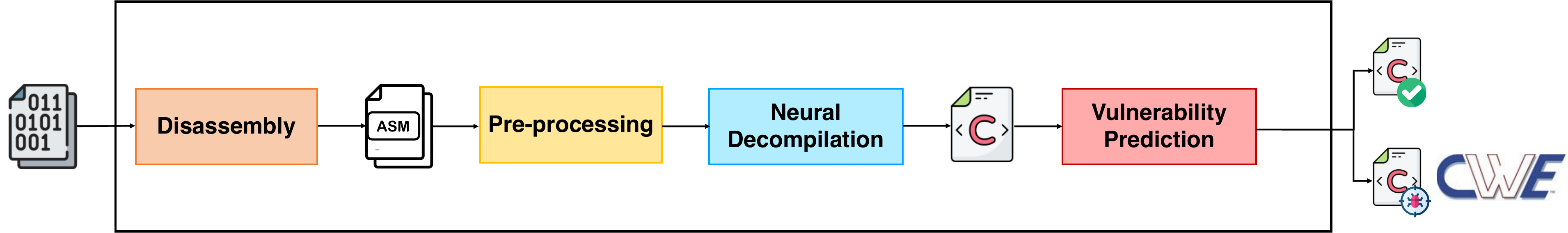}}
\caption{Overview of vulnerability prediction via neural decompilation.}
\label{fig:Solution}
\end{figure*}

In this work, we explore the potential to identify software vulnerabilities in binary files via neural decompilation (Figure \ref{fig:Solution}), without need to access the source code of the target software. 
The results of this study show that: 
\begin{enumerate}
    \item Neural decompilation achieves impressive performance in the decompilation of C/C++ code, reaching up to an Edit Distance of 59\% with the \textit{fairseq} model.
    \item Vulnerability prediction yields highly accurate results:
    \begin{itemize}
        \item the \textit{CodeGPT} model successfully determines whether the code is vulnerable (bi-class analysis, 95\% F1-Score).
        \item the \textit{CodeBERT} model accurately predicts (83\% F1-Score) the type of vulnerability among 20 different vulnerabilities (multi-class analysis), mapped onto the Common Weaknesses Enumeration (CWE) \cite{CWE} defined by MITRE \cite{Mitre}.
    \end{itemize} 
    \item Decompiled source code leads to better vulnerability prediction accuracy compared to disassembled code \cite{schaad2023deep}, showing a 7\% improvement in the bi-class analysis and a 6\% improvement in the multi-class analysis.
\end{enumerate} 

The remainder of the paper is structured as follows. In Section \ref{sec:background}, we elaborate on background concepts about the target problem. In Section \ref{sec:methodology}, we present the methodology for vulnerability prediction via neural decompilation. In Section \ref{sec:experiment}, we illustrate the experimental analysis for the decompilation and vulnerability prediction tasks.  In Section \ref{sec:related}, we examine related work. Section \ref{sec:conclusion} concludes the paper.


\section{Background}
\label{sec:background}

The main complexities of this work lie in the \textit{decompilation} process, i.e., the transformation of executable code into source code. This issue stems from the \textit{semantic gap} between high- and low-level code: machine code does not provide function-level abstraction since function calls are implemented through \textit{call labels} and \textit{jumps} that allow the execution of loops and conditional constructs. This asymmetry causes the loss of the control flow of the original program, which is quite challenging to re-construct from the binary. Several decompilers are available in this field (e.g., Ghidra \cite{Ghidra}, RetDec \cite{RetDec}), achieving reasonable levels of accuracy; however, they come with high costs in terms of knowledge of the specific languages, generation of control flow graphs, and data flow graphs. 
Another complexity in dealing with the translation and vulnerability prediction tasks lies in finding an appropriate dataset. The dataset should meet the following requirements:
\begin{itemize}
    \item \textit{Compilable}: the programs under analysis in the dataset must be compilable to generate executables for the binary analysis.
    \item \textit{Standardized classification}: the programs must be labeled according to the vulnerability they represent following an appropriate categorization system, i.e., a standardized vulnerability taxonomy.
\end{itemize}
    
\begin{table}[htbp]
\caption{C/C++ datasets for vulnerability prediction.}
\begin{center}
\resizebox{\columnwidth}{!}{%
\begin{tabular}{l|ccc}
\toprule
\multicolumn{1}{c}{\textbf{Dataset}} & \textbf{\# Samples} & \textbf{Compilable} & \textbf{Standardized Taxonomy}\\
\midrule
ReVeal \cite{arewethereyet} & $18,169$ & No &  N/A\\
Devign \cite{zhou2019devign} & $26,037$ & No &  N/A\\
Juliet \cite{Juliet} & $128,198$ & \textbf{Yes} & \textbf{Yes}\\
BigVul \cite{bigvul}  & $264,919$  & No  & Only for 8,783 samples\\
DiverseVul \cite{chen2023diversevul}  & 330,492  & No & Only for 16,109 samples\\
\bottomrule
\end{tabular}
}
\end{center}
\label{tab:Datasets}
\end{table}

As shown in Table \ref{tab:Datasets}, among the C/C++ datasets for vulnerability prediction available in the literature \cite{chen2023diversevul, bigvul, arewethereyet, zhou2019devign}, the only dataset suitable for our study is \textit{Juliet C/C++ 1.3} \cite{Juliet}, a synthetic collection of C/C++ programs from the Software Assurance Reference Dataset (SARD) \cite{SARD}, developed by NIST \cite{NIST}. The Juliet test suite was first released in 2010, gaining increasing popularity for benchmarking static analysis tools and vulnerability prediction tasks \cite{zheng2019recurrent, schaad2023deep, li2021vuldeelocator, li2018vuldeepecker, jain2023code, gao2018comprehensive} since it includes both vulnerable and non-vulnerable programs. The dataset provides $64,099$ \textit{compilable} test cases for 118 CWEs. For each CWE, a \textit{test case} comprises a program vulnerable to the target CWE and a non-vulnerable one.

\section{Methodology}
\label{sec:methodology}
\begin{figure*}[!htbp]
\centering
\centerline{\includegraphics[width=0.75\textwidth]{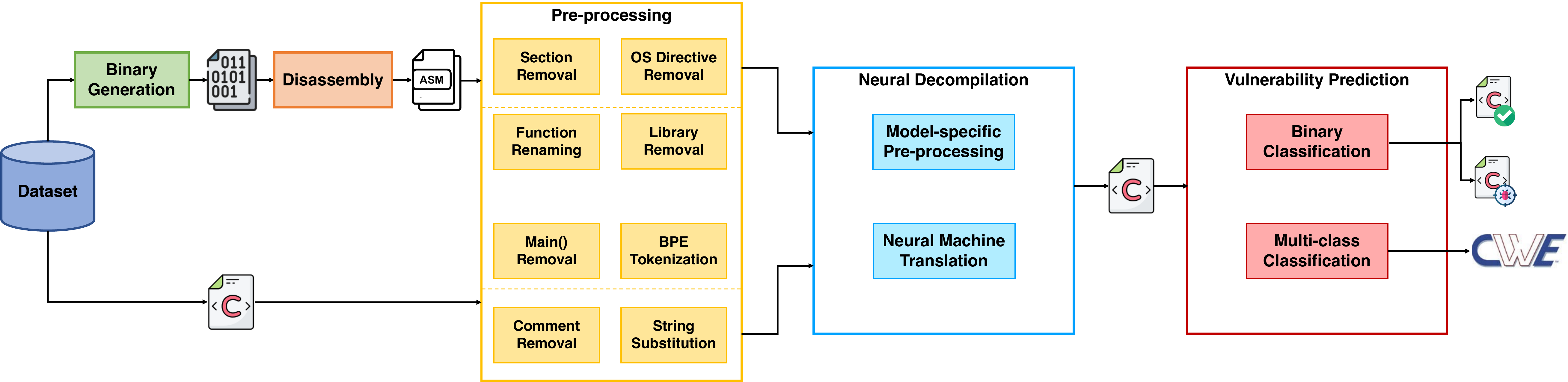}}
\caption{Experimental approach for our study.}
\label{fig:Methodology}
\end{figure*}

Before presenting the experimental approach, we introduce the following assumptions:
 \begin{itemize}
    \item To avoid translating binary code to high-level code directly, we \textit{disassembled} the binary into \textbf{assembly} code. Working directly with binary code would prevent information pre-processing. Consequently, the input for the decompilation model will be assembly code, further pre-processed in the next steps.
    \item Assembly code is defined by the architecture used to generate the executables. In our experimental evaluation, the assembly language is for the \textbf{x86} architecture. 
    \item The high-level target language of the translation is \textbf{C/C++}. These languages are still widely used to develop network applications, Internet of Things (IoT) systems, and systems with safety requirements.
    \item The neural decompilation model must learn to translate code at the \textbf{function level}. Through pre-processing, we strip the programs of unnecessary information associated with libraries, namespaces, external modules, or main functions that only invoke other functions.
\end{itemize}

Our experimental approach revolves around five phases, as shown in Figure \ref{fig:Methodology}. The first phase is \textit{Binary Generation} to compile the C/C++ programs under analysis. We adopted the automated building approach by Richardson \emph{et al.} \cite{JulietCompilation}, configuring the \texttt{gcc/g++} compiler (depending on the program's language) with the \texttt{-O0} optimization level. Due to compilation issues, the dataset was reduced from 128,198 programs to 83,520. The second one is \textit{Disassembly}, which lifts the binary code to assembly code via \textit{objdump} \cite{objdump}. The format of the executables is \texttt{elf64-x86-64}. The third phase is \textit{Pre-processing} of the low- and high-level code. In the low-level code, we removed all sections except \texttt{.text} since we are only interested in the algorithmic logic. We removed library functions defined within the same section, operating system directives, stack initialization, and the \texttt{main} function, which is constant in every program, making its contribution to learning negligible. In the high-level code, we removed the \texttt{main} function, library inclusions, namespaces, comments, and printed strings since there is no solution for the model to predict them. From both low- and high-level code, we replaced function names with standard names (e.g., \texttt{funct0}) to let the model focus exclusively on the algorithm of the actual functions. Finally, we applied toss reduction on tokenized code via BPE, removing programs above the 95th percentile and below the 5th percentile. This reduced the dataset from 83,520 to 75,149 elements. The fourth and fifth phases are \textit{Neural Decompilation} and \textit{Vulnerability Prediction}.

\subsection{Neural Decompilation}
\label{sec:translation}
\textit{Neural decompilation} is an emerging approach to translating binary code into source code. The main advantages of this approach are two. The first is that the decompiled program control-flow structure traces that of the original high-level program, making the results human-readable. This happens because the transformer learns the typical structure of source programs.
The second advantage is that neural decompilation is a data-driven approach, hence it can be trained on new target languages or architectures with new data without depending on specific and hard-to-use toolchains. 
Neural decompilation leverages \textit{Neural Machine Translation} (NMT), a machine learning approach to translating a given sequence of elements into another sequence through the use of \textit{Large Language Models} (LLMs), deep learning algorithms able to perform a set of language processing tasks such as translation, classification or generation. In light of the complexity of the tasks, LLMs generally require large accurate datasets to be trained. The core of the LLMs is the \textit{Transformer} neural network, a deep learning architecture consisting of an encoder-decoder structure based on the multi-headed attention mechanism \cite{vaswani2017attention}, often used for translation from one language to another. In our case, we leverage NMT to translate machine code to high-level code. The fine-tuned models we used for the translation task are \textit{CodeBERT} \cite{feng2020codebert}, \textit{CodeT5+} \cite{wang2023codet5} and \textit{fairseq} \cite{fairseq}. For CodeBERT, we combined it with a decoder to have an encoder-decoder architecture suitable for code generation \cite{wang2021codet5}.

\subsection{Vulnerability Prediction}
\label{sec:classification}
This stage represents the core of the \textit{vulnerability prediction} task. The goal is to identify whether a target program is \textbf{vulnerable}, along with the \textbf{type of vulnerability} affecting it. We addressed these problems as \textit{classification problems}, following the direction of several studies in the field \cite{zheng2019recurrent, schaad2023deep, pewny2015cross}.
Specifically, we faced the following classification tasks:
\begin{itemize}
    \item \textit{Binary classification}: deals with finding whether a target C/C++ function is vulnerable (\textit{bad}) or not (\textit{good}), regardless of the specific vulnerability.
    \item \textit{Multi-class classification}: aims to identify which vulnerability affects the program. Vulnerability types are the classes that the classification models predict. Such vulnerabilities are mapped to the CWE \cite{CWE}.
\end{itemize}
Therefore, we train several models on a training set consisting of original C/C++ programs, i.e., non-decompiled, and then test them on a test set consisting of neurally decompiled C/C++ programs. 
Looking at previous studies on this topic \cite{chen2023diversevul, schaad2023deep, zheng2019recurrent}, we selected models that can handle the complexity of vulnerability prediction, including \textit{CodeBERT} \cite{feng2020codebert}, \textit{CodeT5+} \cite{wang2023codet5}, \textit{CodeGPT} \cite{huggingface2020codegpt}, \textit{Simple Recurrent Neural Networks (SRNN)}, \textit{Long Short-Term Memory (LSTM)} and \textit{Gated Recurrent Unit (GRU)}.

\begin{table*}[!htbp]
\caption{Automatic evaluation of the neural decompilation task. Best performance is \textcolor{blue}{\textbf{blue}}.}
\begin{center}
\resizebox{0.7\textwidth}{!}{%
\begin{tabular}{cc|cc|cccc}
    \toprule
    \textbf{Approach} & \textbf{Model} & \textbf{\# functions} & \textbf{Max target length} & \textbf{BLEU-4} & \textbf{ED} & \textbf{METEOR} & \textbf{ROUGE-L}\\
    \midrule
    \textit{*Ghidra} \cite{Ghidra} &  \textit{IR} & 1,000 & 338 & 22\% \DrawPercentageBar{0.22} & 23\% \DrawPercentageBar{0.23} & 29\% \DrawPercentageBar{0.29} & 22\% \DrawPercentageBar{0.22} \\ 
    °Katz \emph{et al.}  \cite{katz2018using} &  \textit{RNN} & 700,000 & 88 & -  & 30\% \DrawPercentageBar{0.30} & -  & -   \\ 
    °Hosseini \emph{et al.} \cite{hosseini2022beyond} &  \textit{fairseq} & 2,000,000 & 271 & -  & 54\% \DrawPercentageBar{0.54} & -  & -   \\  
    \midrule

    \multirow{3}{*}{\textit{Proposed approach}}
    
    & \textit{CodeBERT}  & 75,149 & 338 & 31\% \DrawPercentageBar{0.31}  & 39\% \DrawPercentageBar{0.59} & 43\% \DrawPercentageBar{0.43}  & 56\% \DrawPercentageBar{0.56} \\
    & \textit{CodeT5+} & 75,149 & 338 & 48\% \DrawPercentageBar{0.48}  & 49\% \DrawPercentageBar{0.49} & 55\% \DrawPercentageBar{0.55} & 70\% \DrawPercentageBar{0.70} \\
    & \textit{fairseq} & 75,149 & 338 & \textcolor{blue}{\textbf{58\%}} \DrawPercentageBarBlue{0.58} & \textcolor{blue}{\textbf{59\%}} \DrawPercentageBarBlue{0.59} & \textcolor{blue}{\textbf{64\%}} \DrawPercentageBarBlue{0.64} &  \textcolor{blue}{\textbf{77\%}} \DrawPercentageBarBlue{0.77} \\   
    \bottomrule 
\end{tabular}
}%
\label{tab:Neural decompilation evaluation}
\end{center}
\centering
\footnotesize*: results obtained on a random test set of Juliet dataset. \hspace{1.5cm} °: trained with different data than Juliet.
\normalsize
\end{table*}

\section{Experimental Analysis}
\label{sec:experiment}

\subsection{Evaluation - Neural Decompilation}
\label{sec:exp_eval_decompilation}
We evaluated the translation quality using output similarity metrics including BLEU, METEOR, ROUGE, and Edit Distance (ED), illustrated in Table \ref{tab:Neural decompilation evaluation}. Each metric expresses different characteristics of translation quality.
One of the most relevant metrics in code translation is ED, suitable for capturing small variations in the code and representative of the number of changes to perform on the decompiled output. We refer to this metric as ED for consistency with prior work, but it represents a normalized ED measure (Edit Similarity) ranging from 0, the worst case, to 1, the best case. 

\begin{table*}[!ht]
\caption{Qualitative analysis on decompilation.}
\label{tab:Translation Examples}
\footnotesize
\centering
\resizebox{0.75\textwidth}{!}{%
\begin{tabular}{
 >{\arraybackslash}m{5.3cm} |
 >{\arraybackslash}m{5.4cm} |
 >{\arraybackslash}m{5.7cm}}
\toprule
\textbf{Ground Truth} & \textbf{Neural Decompilation} & \textbf{Ghidra}\\ \midrule
\texttt{\textcolor{blue}{static void} funct0(\textcolor{blue}{long} *data) \{}  & \texttt{\textcolor{blue}{static char}* funct0(\textcolor{blue}{char} *data)\{ }       & \texttt{\textcolor{blue}{void} funct0(\textcolor{blue}{long} *param\_1)\{} \\
\hspace{0.5cm}\texttt{\textcolor{blue}{delete} data;}         & \hspace{0.5cm}\texttt{data = \textcolor{blue}{new char}[100]; }     & \hspace{0.5cm}\texttt{\textcolor{blue}{if} (param\_1!=(\textcolor{blue}{long} *)0x0) \{} \\
\texttt{\}}                                 & \hspace{0.5cm}\texttt{\textcolor{blue}{return} data;}               & \hspace{0.7cm}\texttt{\textcolor{blue}{operator.delete}(param\_1,8);\}} \\
\texttt{\textcolor{blue}{void} funct1()\{ }                   & \texttt{\}}                                       & \hspace{0.5cm}\texttt{\textcolor{blue}{return}; }                     \\
\hspace{0.5cm}\texttt{\textcolor{blue}{long} * data;}         & \texttt{\textcolor{blue}{void} funct1()\{}                          & \texttt{\}}           \\
\hspace{0.5cm}\texttt{data = NULL;}         & \hspace{0.5cm}\texttt{\textcolor{blue}{char} * data ;}              & \texttt{\textcolor{blue}{void} funct1(\textcolor{blue}{void}) \{}                  \\
\hspace{0.5cm}\texttt{data = \textcolor{blue}{new} \textcolor{blue}{long}[100];}& \hspace{0.5cm}\texttt{data = NULL ;}              & \hspace{0.5cm}\texttt{\textcolor{blue}{long} *plVar1;}\\
\hspace{0.5cm}\texttt{funct0(data);}        & \hspace{0.5cm}\texttt{data = funct0(data); }      & \hspace{0.5cm}\texttt{plVar1 = (\textcolor{blue}{long} *)\textcolor{blue}{operator.new}[](800);}  \\
\texttt{\}}                                 & \hspace{0.5cm}\texttt{\textcolor{blue}{delete} data ;}              & \hspace{0.5cm}\texttt{funct0(plVar1);} \\
                                            & \texttt{\}}                                       & \hspace{0.5cm}\texttt{\textcolor{blue}{return}; } \\
                                            &                                                   & \texttt{\}} 
\\ \midrule
\textbf{ED Score} & 57\% \DrawPercentageBar{0.57} & 37\% \DrawPercentageBar{0.37}\\ \bottomrule
\end{tabular}
}
\end{table*}

Table \ref{tab:Neural decompilation evaluation} compares different decompilation solutions based on their approach, number of functions used for training and testing, maximum target length, and output similarity metrics.
The results of neural decompilation are outstanding. Specifically, fairseq achieves a 59\% ED score using $75,149$ C/C++ functions for both training and testing with a maximum target length of 338 tokens. These results are close to those of other studies on this topic: Hosseini \emph{et al.} \cite{hosseini2022beyond} reached an 54\% ED using about $2,000,000$ C/C++ functions with a maximum length of $271$, while Katz \emph{et al.} \cite{katz2018using} achieves an ED of 30\% using about $700,000$ C-code snippets with a maximum target length of $88$. It is worth noting that both studies leverage a number of samples at least an order of magnitude higher than ours. In addition, the function lengths analyzed with our approach are greater than those used by either work. For the sake of clarity, we note that these works adopt a complementary version of ED, less widespread for NMT tasks \cite{liguori2023evaluates}. 
Focusing on the other metrics obtained from fairseq, ROUGE-L shows that, in 77\% of cases, the generated code contains word sequences similar to those of the reference. We observe a 4-gram overlap of 58\% (BLEU-4) and a token order adherence of 64\% (METEOR).

For the sake of completeness, we also evaluated the performance of a traditional decompiler, Ghidra \cite{Ghidra}. Specifically, we randomly sampled 1,000 programs from our decompiled test set. Ghidra relies on control flow graphs (CFGs), data flow graphs (DFGs), and other intermediate representations (IRs) to perform decompilation. Ghidra results, with a 23\% ED, indicate significantly lower performance compared to studies utilizing neural network-based models, which suggests that traditional graph-based methods may struggle to capture the complexities of the source code as effectively as neural approaches. However, unlike a traditional decompiler, an NMT model always generates an answer despite having a low probability of approaching the correct one. In addition, one of the disadvantages of using a neural network is that there is less control over the translation, i.e., in case of errors it is very complex to determine which factor led the model to a specific decision over others. 
\begin{table*}[htbp]
\centering
\caption{Models performance for the binary and multi-class classification task. Best performance is \textbf{\textcolor{blue}{blue}}.}
\label{tab:Classifications}
\resizebox{0.75\textwidth}{!}{%
\begin{tabular}{cc|cc|cc}
\toprule
 &   & \textbf{Binary Classification}  & & \textbf{Multi-class Classification}  & \\ \midrule
\textbf{Approach} & \textbf{Model} & \textbf{Accuracy (\%)} & \textbf{F1-Score (\%)} & \textbf{Accuracy (\%)} & \textbf{F1-Score (\%)}\\ \midrule

\multirow{1}{*}{\textit{Flawfinder*}} \cite{flawfinder}

& \textit{Static Analysis}  & 52 \DrawPercentageBar{0.52} & 49 \DrawPercentageBar{0.49} & 13 \DrawPercentageBar{0.13} & 11 \DrawPercentageBar{0.11}\\

\multirow{1}{*}{Schaad \emph{et al.}°} \cite{schaad2023deep}

& \textit{SRNN} & 88 \DrawPercentageBar{0.88} & 88 \DrawPercentageBar{0.88} & 77 \DrawPercentageBar{0.77} & 79 \DrawPercentageBar{0.79}\\

\midrule

\multirow{6}{*}{\textit{Proposed approach}}

& \textit{SRNN} & 88 
\DrawPercentageBar{0.88} & 88 \DrawPercentageBar{0.88} & 72 \DrawPercentageBar{0.72} & 70 \DrawPercentageBar{0.70}\\ 
& \textit{LSTM} & 94 
\DrawPercentageBar{0.94} & 94 \DrawPercentageBar{0.94} & 75 \DrawPercentageBar{0.75} & 72 \DrawPercentageBar{0.72}\\
& \textit{GRU} & 91 
\DrawPercentageBar{0.90} & 90 \DrawPercentageBar{0.90} & 78 \DrawPercentageBar{0.78} & 77 \DrawPercentageBar{0.77}\\

& \textit{CodeT5+} & 94 \DrawPercentageBar{0.94} & 94 \DrawPercentageBar{0.94} & 81 \DrawPercentageBar{0.81} & 81 \DrawPercentageBar{0.81}\\
& \textit{CodeGPT} & \textcolor{blue}{\textbf{95}} \DrawPercentageBarBlue{0.95} & \textcolor{blue}{\textbf{95}} \DrawPercentageBarBlue{0.95} & 82 \DrawPercentageBar{0.82} & 82 \DrawPercentageBar{0.82}\\
& \textit{CodeBERT} & 94 \DrawPercentageBar{0.94} & 94 \DrawPercentageBar{0.94} & \textcolor{blue}{\textbf{83}} \DrawPercentageBarBlue{0.83} & \textcolor{blue}{\textbf{82}} \DrawPercentageBarBlue{0.82}\\
\bottomrule
\end{tabular}
}%
\\
\vspace{1pt}
\footnotesize*: multi-class concern Flawfinder-compatible CWEs. \hspace{0.5cm}
°: multi-class concern different CWEs selected by Schaad et al. \cite{schaad2023deep}.
\normalsize
\end{table*}
An interesting finding is that pre-trained models, i.e., \textit{CodeT5+} and \textit{CodeBERT}, perform considerably worse than the model trained from scratch, i.e., fairseq. As noted by Tufano \emph{et al.} \cite{tufano2023automating}, we can associate the negative impact of pre-training with the size of the training dataset. Pre-training helps boost performance only if the amount of fine-tuning data available is small. We deem our dataset with about 75,000 samples large enough to overwrite the parameters tuned in the pre-training phase. According to these results, we adopt \textit{fairseq} as the neural decompiler for further analysis. 

\noindent
\textbf{Qualitative analysis}.
Table \ref{tab:Translation Examples} illustrates an example of ASM-C/C++ code translation. In the reported vulnerable example associated with CWE \textit{762 Mismatched Memory Management Routines}, the potential flaw is to deallocate the array \texttt{data} in heap memory with \texttt{delete} instead of \texttt{delete []}. 
Specifically, we report two versions of decompiled code for the example, one produced by our neural decompilation model (fairseq) and the other through Ghidra \cite{Ghidra}. 

On the one hand, it is possible to notice that neural decompilation retraces the original code structure. The resulting code achieves an ED score of 57\%. Although some operations are reversed in the corpus of the two functions, the logical sense and semantics of the code are correct. In addition, the code is extremely readable. However, this is not always the case. 
Sometimes, the translated program leads us to consider that although ED may be high, the logical sense of the program could be incorrect. 

On the other hand, the code produced by Ghidra closely resembles the assembly structure and scores a 37\% ED. It is worth noting that this code is harder to understand for a human operator. Moreover, to perform decompilation, Ghidra needs the entire assembly file, not just the function to decompile. 
However, the key advantage of Ghidra, as a traditional decompiler, is that it offers a higher probability that the decompiled code is functionally and semantically correct, though this is not guaranteed.

\vspace{-5pt}
\begin{table*}[ht!]
\caption{Comparison of vulnerability prediction works on binary files.}
\label{tab:Related work comparison}
\centering
\resizebox{0.8\textwidth}{!}{%
\footnotesize 
{\color{black}\begin{tabular}[h!]{c|ccc|cccc}
\toprule
\normalsize\textbf{Work}  & 
\normalsize\textbf{Dataset} & 
\normalsize\textbf{Classes of} & 
\normalsize\textbf{Standardized} & 
\normalsize\textbf{Intermediate} &
\textbf{\normalsize Vulnerability} &
\multicolumn{1}{c}{\normalsize\textbf{Bi-Classification}} &
\multicolumn{1}{c}{\normalsize\textbf{Multi-Classification}} \\
&  & \normalsize\textbf{Vulnerabilities}  & \normalsize\textbf{Taxonomy} & \textbf{\normalsize Representation} & \textbf{\normalsize Predictor} & \textbf{\normalsize Accuracy} & \textbf{\normalsize Accuracy}\\


\toprule

Pewny \emph{et al.} \cite{pewny2015cross} & 
    OpenSSL
    &
    5
    &
    \normalsize\ding{55}
    &
    IR+CFG 
    & 
    Greedy
    &
    \normalsize\ding{55}
    &
    61\% \DrawPercentageBar{0.61}
    \\ \midrule

\emph{DiscovRE} \cite{eschweiler2016discovre} & 
    Github
    &
    2
    &
    \normalsize\ding{55}
    &
    CFG
    &
    k-NN
    &
    \normalsize\ding{55}
    &
    93\% \DrawPercentageBar{0.93}

    \\ \midrule

Dahl \emph{et al.} \cite{dahl2020stack} & 
    Open-source Code
    &
    1
    &
    \normalsize\ding{55}
    &
    ASM Tokens
    &
    RNN
    &
    \normalsize\ding{55}
    &
    99\% \DrawPercentageBar{0.99}
    \\ \midrule

\emph{VDISCOVER} \cite{grieco2016toward} & 
    Debian
    &
    4
    &
    \normalsize\ding{55}
    &
    Call Graph
    &
    Random Forest
    &
    \normalsize\ding{55}
    &
    55\% \DrawPercentageBar{0.55}

    \\ \midrule
    
Zheng \emph{et al.} \cite{zheng2019recurrent} & 
    SARD
    &
    4
    &
    CWE
    &
    LLVM IR
    & 
    RNN
    &
    \normalsize\ding{55}
    &
    90\% \DrawPercentageBar{0.9} 
 
    \\ \midrule

Schaad \emph{et al.} \cite{schaad2023deep} & 
    Juliet
    &
    23 
    &
    CWE 
    &
    LLVM IR
    &  
    RNN
    &
    88\% \DrawPercentageBar{0.88}
    &
    77\% \DrawPercentageBar{0.77} 
    \\ \midrule\midrule

\textbf{Proposed solution} & 
    \textbf{Juliet}
    &
    \textbf{20}
    &
    \textbf{CWE}
    &
    \textbf{Decompiled C/C++}
    &
    \textbf{Transformers}
    &
    \textbf{95\%} \DrawPercentageBar{0.95}
    &
    \textbf{83\%} \DrawPercentageBar{0.83}
    \\ \bottomrule
\end{tabular}}
}%

\end{table*}
\subsection{Evaluation - Vulnerability Prediction}
\label{sec:exp_eval_vuln_pred}

\noindent
\textbf{Binary Classification}.
\label{sec:binary}
The first analysis concerns the classification of the target code as vulnerable/non-vulnerable. We submitted the C/C++ code translated by the neural decompilation model to classification, adopting the same data pre-processing for each model. We split the data 80-20\%,  ensuring stratification by target and maintaining a 50-50\% balance between vulnerable and non-vulnerable samples. We selected \textit{accuracy} and \textit{F1-Score} as the metrics to evaluate the performance of the models. 

Table \ref{tab:Classifications} shows the models' performance for the binary classification on the decompiled C/C++ code. CodeGPT is the best-performing model with an accuracy and F1-Score of 95\%. 
The highest gap lies between SRNN and CodeGPT, exhibiting a discrepancy of about 7\% in terms of F1-Score. Table \ref{tab:Classifications} also reports the performance of a state-of-the-art work on vulnerability prediction in binary files \cite{schaad2023deep}, which achieves an 88\% F1-Score through the LLVM IR and the SRNN model, further described in Table \ref{tab:Related work comparison} and Section \ref{sec:related}.

Furthermore, we tested a static analysis tool, Flawfinder \cite{flawfinder}, on the same test set. This tool scans C/C++ source code and flags potential security holes. Flawfinder exhibits metrics that are close to 50\% on average, which is equivalent to a random model. Therefore, the results support the implemented design choice of using DL models for vulnerability prediction instead of a static analysis tool.
Finally, we leveraged \textit{Shap} \cite{NIPS2017_7062}, a tool for explaining the output of our vulnerability prediction models, to better understand which features are the most influential for the learning process, i.e., the tokens most exploited for classification. This is a helpful analysis to avoid falling into the so-called \textit{Clever Hans effect}, which occurs when the model produces correct predictions based on "wrong" features \cite{kauffmann2020clever}. The results revealed tokens in C/C++ code such as: \textit{if}, \textit{new}, \textit{strcat}, \textit{funct}, \textit{static}, \textit{malloc}, and \textit{source}. These results allow us to state that the model focuses on tokens that are relevant to implementing countermeasures to vulnerabilities, e.g., \textit{if} constructs to check specific conditions (e.g., underflow or overflow).

\noindent
\textbf{Multi-class Classification}.
\label{sec:multi-class}
This analysis aims to identify the CWE associated with a vulnerable function. Specifically, the number of classes (CWEs) for this classification task is 20 (190, 762, 121, 590, 194, 197, 78, 36, 23, 690, 369, 457, 134, 680, 758, 400, 789, 253, 398, 761), plus an additional class associated with non-vulnerable (good) functions. We obtain a total of 42,483 samples, with 20,393 corresponding to vulnerable functions and 22,090 to non-vulnerable ones. The complexity of this task stems from similarities between test cases across classes. Rather than selecting classes solely by frequency, we manually chose those with distinct CWEs to minimize model confusion with similar ones. Consequently, to the best of our knowledge, we evaluated the difference between CWEs from a logical point of view.
We prioritized the CWEs that belonged to the Top 25 CWEs published yearly by MITRE among the most represented ones, i.e., those with a reasonable amount of samples. Finally, we split the data 80\%-20\%, ensuring stratification by CWE.

Table \ref{tab:Classifications} shows the accuracy and F1-Score of the models for the multi-class classification task. The models' performance decreased from the binary classification case, with CodeBERT performing the best with an accuracy of 83\% and F1-Score of 82\%. The performance gap between our approach, utilizing CodeBERT, and the method proposed by Schaad \emph{et al.} \cite{schaad2023deep} is approximately 6\% in terms of accuracy and 3\% in the F1-score. However, it is important to note that the 20 classes analyzed in our study differ both in type and quantity from their 23 classes \cite{schaad2023deep}, as shown in  Table \ref{tab:Related work comparison}. 

Flawfinder \cite{flawfinder} is compatible with only a subset of CWEs \cite{CWEflawfinder}. Consequently, we conducted an experiment excluding all non-compatible CWEs from our dataset, resulting in approximately 500 programs. The performance in this test was notably worse compared to our earlier results on the binary classification task, highlighting the potential advantage that deep learning techniques could offer in this context.

Table \ref{tab:Related work comparison} highlights the key improvements of the proposed approach in its entirety compared to similar works. Notably, in addition to achieving higher performance, our method uniquely leverages the transformer architecture for both tasks and includes source code as an intermediate representation. Therefore, we can conclude that relying on source code as IR, instead of disassembled code, resulted in a 7\% increase in bi-class analysis performance and a 6\% improvement in multi-class.

\section{Related Work}
\label{sec:related}
Table \ref{tab:Related work comparison} summarizes the comparison between the related work and our solution.
Our work differs from the others, in terms of binary code analysis and vulnerability prediction. On the one hand, neural decompilation does not need any formal representation of the binary code (e.g., control flows, intermediate representation) since the code is treated as plain text for the decompilation task.  On the other hand, contrary to other works that focus on technology-specific or single vulnerabilities, we focus on identifying several vulnerabilities described in standard taxonomies, namely the CWE provided by MITRE, achieving higher accuracy than two related works \cite{pewny2015cross,grieco2016toward} and also surpasses the only work that encompasses a comparable number of vulnerabilities \cite{schaad2023deep}. Even though the other works achieve a slightly better performance, they only analyze a significantly smaller number of technology-specific vulnerabilities \cite{zheng2019recurrent, eschweiler2016discovre, dahl2020stack}. Using a standardized vulnerability taxonomy is a key aspect in evaluating these works: technology-specific vulnerabilities are harder to generalize and typically do not apply to different systems, while standardized vulnerabilities, e.g., CWEs, represent global definitions that span several kinds of software systems and products.

\section{Conclusion}
\label{sec:conclusion}
In this work, we presented an experimental study on vulnerability prediction in binary files. In particular, we leveraged neural decompilation to translate binary code into source code, treating both as simple text. Our goal was to demonstrate that a more effective solution for detecting vulnerabilities in binary files is to use neurally decompiled source code, rather than relying on complex toolchains to convert binary code into an intermediate representation or to extract control flows for decompilation.
Our experimental analysis revealed that neural decompilation is fit for this task, achieving great results comparable with state-of-the-art works, and that transformer-based and DL models can accurately predict whether a binary file is vulnerable and identify the type of vulnerabilities, mapped to MITRE CWE, using the neurally decompiled code.

\section*{Acknowledgments}
\label{sec:Ack}
This work has been partially supported by the PNRR - DM 118/2023 project, \textit{Mis.: I.3.4 Dottorati Transizione Digitale} (CUP E66E23001030002), and the \textit{GENIO} project, funded by MIMIT (CUP B69J23005770005).


\bibliographystyle{IEEEtran}
\bibliography{biblio}

\end{document}